# Ionic Charge Distributions in Silicon Atomic Wires

Jeremiah Croshaw[1,2], Taleana Huff[3], Mohamad Rashidi[1], John Wood[1], Erika Lloyd[1], Jason Pitters[3], Robert Wolkow[1,2,3]

[1]Department of Physics, University of Alberta, Edmonton, Alberta, T6G 2J1, Canada

[2]Quantum Silicon Inc., Edmonton, Alberta, T6G 2M9, Canada

[3]Nanotechnology Research Centre, National Research Council Canada, Edmonton, Alberta, T6G 2M9, Canada

**Abstract**

Using a non-contact atomic force microscope (nc-AFM), we examine continuous DB wire structures on the hydrogen terminated silicon (100) 2x1 surface. By probing the DB structures at varying energies, we identify the formation of previously unobserved ionic charge distributions correlated to the net charge of DB wires and their predicted lattice distortion. Performing spectroscopic analysis, we identify higher energy configurations corresponding to alternative lattice distortions as well as tip-induced charging effects. By varying the length and orientation of these DB structures, we further highlight key features in the formation of these ionic surface phases.

The development of novel "beyond CMOS" devices has led to the design and implementation of device components made of few to single atoms [1–8]. Due to their decreased size, the charge distribution of such devices can no longer be treated as an ensemble average, instead requiring discrete addressment. The atomic force microscope (AFM) has been proven to be a useful tool for the detection of discrete transitions between charge states in both molecular [9–13] and atomic structures [14–16], including silicon dangling bonds (DBs) [6,17–19]. Enabled through atomically precise lithography [20–25], fabricated DB systems are capable of representing binary logic through charge state manipulation within many-dot structures [6,17–19] . Not elucidated in these studies was a thorough investigation of continuous 1-D DB wires which are predicted to exhibit either a spin or ionic ordering [26–29] similar to that seen in low dimensional Mott insulators [30–32].

In this work we further the understanding of continuous DB structures at previously unexplored energies within the bulk Si band gap as enabled by AFM. Due to the anisotropy of the (2x1) reconstruction [33–36], these DB structures are either patterned along a single dimer, creating what is commonly referred to as a bare dimer [37–40], or along the same side of a dimer row creating a DB wire [41,42]. DBs within a bare dimer are separated only by the shared dimer bond that forms a pi bond [26,43–45], while DBs within a DB wire are separated by a second layer back bonded Si atom. This difference in bonding leads to different constraints in regard to lattice relaxation, motivating differences in their atomic and electronic properties. By using a charge sensitive Si-functionalized probe [25,46], we are able to clearly distinguish between the positive, neutral, and negative charge states of DBs within bare dimers and DB wires, yielding insights into the association of DB charging with lattice distortion.

We identify the formation of an ionic charge distribution within bare dimers and DB wires, with the DB wires only exhibiting ionic character when accompanied by changes in their overall charge state. Through spectroscopic analysis, we identify the formation of both higher energy ionic configurations attributed to an alternative lattice distortion [27] as well as higher order charge distributions within DB wires, with differences observed for DB wires of varying lengths.

DBs on the otherwise hydrogen terminated Silicon (H-Si) (100)-2x1 surface have been shown to behave as quantum dots capable of holding 0, 1, or 2 electrons (rendering the DB in a positive, neutral, or negative charge state) [33,34,47,48]. The native charge state of a single DB can be modified by either varying the crystal doping level or the electrostatic environment surrounding the DB [6,17]. Using a degenerately n-doped crystal (see methods), Kelvin probe force microscopy (KPFM or Δf(V) spectroscopy) reveals the charge transitions of the DB through discrete shifts in Δf. The transition bias between the neutral and negative (0/-) and the positive and neutral (+/0) charge states are routinely observed 0.2-0.4 V below the Fermi-level [6,17,18] and at the onset of the valence band [34,49], respectively. Due to the location of the (+/0) charge transition, high tunneling current appears when probing it with KPFM in the expected bias range, making it challenging to maintain tip integrity and distinguish the DB charging from higher bulk current contributions [50]. By raising the electrostatic potential surrounding the DB with local fixed charges [17], these charge transitions can be shifted in energy to a mid-gap region allowing for both charge transitions to be detected with KPFM. Figure 1a shows a Δf(V) spectroscopy of a DB whose position on the edge of a 14 DB cluster (shown in the supporting information in Figure S1) caused its charge transitions to shift more mid gap—the positive (purple), neutral (orange), and negative (blue) charge state are highlighted. The (+/0) charge transition is seen at $V_S$ = -0.8 V and the (0/-) charge transition at $V_S$ = 0.7 – 0.8 V, which is ~1.2 V higher than the (0/-) values previously reported [6,17,18] (corresponding I(V) spectroscopy and STM images can be seen in Figure S2). The slanted nature of the KPFM charge step is not seen in other reported single electron charging systems [11,51] and is thought to be a result of a current-induced averaging of the charge state based on competing emptying and filling rates of the DB [49,50].

Constant height Δf maps of the DB in a fixed positive (+), neutral (0), and negative (-) charge state relative to the H-Si surface can be seen in Figure 1 (b),(c), and (d), respectively. Matching the relative frequency shifts seen in the Δf(V) spectroscopy, the (+) DB appears "brighter" than the surrounding H-Si atoms, the (0) DB shows similar contrast to the surrounding H-Si atoms, and the (-) DB appears as "darker" than the surrounding H-Si atoms. In addition to the change in charge state, there is also a predicted corresponding lattice distortion of the host Si atom [18,47,52]. A (-) DB raises a host Si atom up (u) from the surface by roughly 30 pm as it adapts more $sp^3$ character while a (+) DB lowers the position of the host Si atom down (d) towards the surface by roughly 40 pm with it adapting more $sp^2$ character [47]. This lattice distortion is qualitatively shown in the ball-and-stick models in Figures 1 (b)-(d) with the horizontal black line showing the relative shift from the (0) Si atom in grey. It is difficult to extract such lattice features from AFM data as the measured Δf signal is convoluted with the changing covalent and charge induced dipole interaction of the surface DB with the neutral DB of the Si terminated tip [53–57].

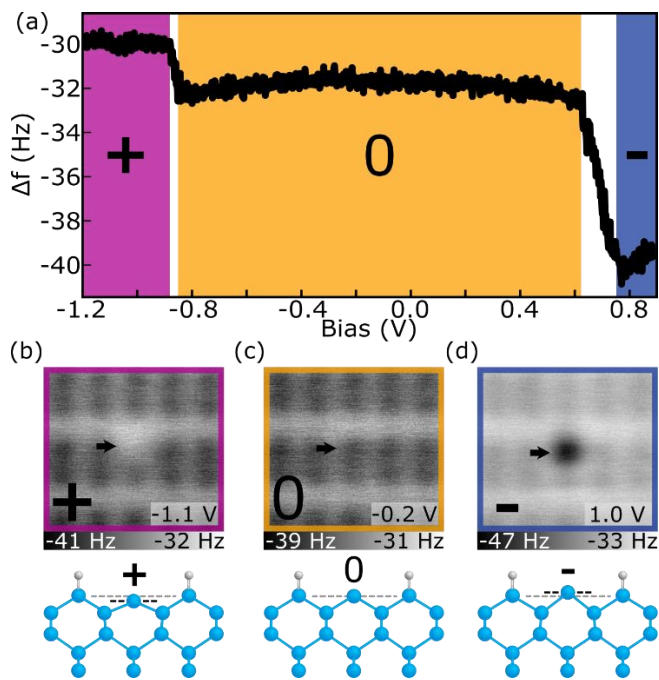

Figure 1. **Three Charge States of the Si DB.** (a) Δf(V) spectroscopy over a Si DB showing the (+/0) and (0/-) charge transitions (occurring at ~ -0.9 V and ~0.7-0.8 V, respectively). (b),(c),(d) Constant height Δf images of the DB in a fixed positive (+), neutral (0), and negative (-) charge state, respectively, with corresponding qualitative ball and stick models of the Si DB in each charge state. The black dashed line in the models highlights the vertical shift in height of the charged host Si atom relative to the neutral state (grey dashed line), with dimensions listed in the text. Each image is 2.1 x 2.1 nm$^2$ with bias values indicated in the lower right of the panels. All measurements performed at -300 pm from the height set point (see methods).

Consistent with observations of the unterminated 2x1 surface [58–64], bare dimers are predicted to undergo a lattice distortion putting one atom in a raised (sp$^3$) electron-rich (-) state, and the other in a lowered (sp$^2$) electron-deficient (+) state [26,65,66]. Figure 2a shows the Δf(V) spectroscopy of each DB in a bare dimer (labelled left and right) revealing six distinct regions. Δf(V) line scan maps along the bare dimer were also taken, with 50 line scans over the dimer acquired at incremented bias intervals of 0.02 V (corresponding I(V) spectroscopy and I(V) line scan maps are shown in Figure S3).

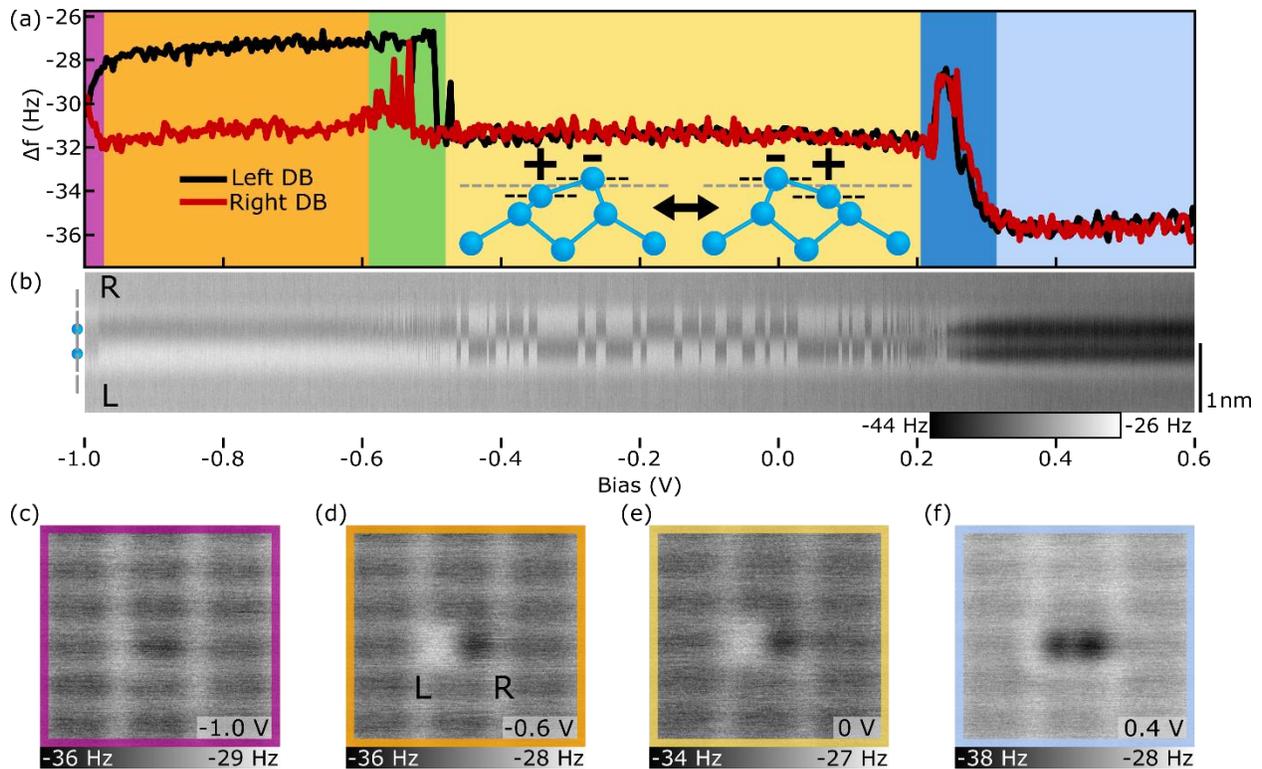

Figure 2. **Charge Distributions of the Bare Dimer.** (a) Δf(V) spectroscopy over the right (red) and left (black) DB of the bare dimer. The inset shows the buckled geometry of the bare dimer in the two degenerate charge configurations (+-/du or -+/ud) (dimensions taken from ref [67]). (b) Δf(V) line scans over the bare dimer with the position of the bare dimer within the line scans highlighted by the model on the left of (b) and a scale bar of 1 nm on the right. 50 line scans from a bias of -1.0 to 0.6V are shown, in increments of 0.02 V. (c)-(f) Constant height Δf images of the bare dimer in four of the six bias regions of (a), as indicated by the coloured border and bias. All measurements were performed at a tip-sample distance -300 pm from the height set point (see methods). Corresponding band diagrams are shown in Figure S7.

At biases lower than -0.98 V (purple), the bare dimer appears to be in a symmetric configuration with both sides of the bare dimer appearing neutral also shown in the constant height Δf image of Figure 2c. Increasing the sample bias to between -0.98 V and -0.6 V (orange), the bare dimer reorders to the expected ionic (+-,down-up) configuration, with the left DB appearing positive (less negative Δf ) and the right DB appearing negative (more negative Δf), as supported in Figure 2d. Between -0.6 V to -0.5 V (green), the tip becomes resonant with the pi state [26,43–45] of the bare dimer, possibly allowing charge to quickly transfer between sides of the bare dimer as indicated by the fluctuations in signal in the Δf(V) spectroscopy of Figure 2a, as well as the streakiness seen in the line scans of Figure 2b. Between -0.5 V and 0.2 V (yellow), the Δf(V) spectroscopies show that both DBs appear negative, however looking at the line scans in Figure 2b and the Δf map of Figure 2e, it is clear that the bare dimer is still in an ionic configuration, except now it readily switches between its (-+) and (+-) configurations, likely facilitated by the newly accessible pi state. The reason both DBs in the Δf(V) spectroscopies appear negative is due to a localized attractive interaction between the DB and the tip causing the negative charge within the bare dimer to "follow" the tip [18,55,59]. Thus, a DB within a bare dimer that is measured with the tip overhead for a sufficient time interval always appears negative. Such an effect, however, is not seen in the line scans of Figure 2b, likely due to the speed of the scans; the tip is not localized over a single DB for a long enough time for a tip induced charge redistribution to occur.

At higher bias values between 0.2 V and 0.3 V (dark blue), the bare dimer becomes resonant with the pi* state [44,45] as indicated by the peaks seen in the Δf(V) spectroscopy. Looking at the line scans in Figure 2b, it appears that the DBs experience this increase in Δf simultaneously suggesting that this peak is a measurement artifact, i.e, electron dynamics are occurring at a faster rate than the AFM is capable of sampling. Corresponding features are also seen in the I(V) plots in Figure S3, corresponding to resonant tunneling through the pi* state. Above bias values of 0.3 V (light blue), both DBs in the bare dimer appear to be in a negative charge state as supported by Figure 2f and the more negative Δf signal seen for both DBs in Figure 2a. It is unclear if the bare dimer is in a 2e$^-$ (--) charge configuration due to electron injection from the tip, or is switching between a degenerate 1e$^-$ (0-/-0) charge configuration at rates faster than the sampling of the AFM. Since the origin of the characteristic butterfly shape [38,39] of the bare dimer in STM images (Figure S3) is attributed to a flipping of between two degenerate states [38], it is more likely that the bare dimer exists in a 1e$^-$ (0-)/(-0) configuration. Further discussions regarding the structural and charge distributions of the bare dimer are given in the Supporting Information.

DB wires are predicted to undergo a similar lattice distortion attributed to a Peierls effect [68–70], thus reducing the overall energy of the DB wire. Features in STM measurements have been attributed to such reorderings through measured variations in the height of DBs within a wire, as explained by Jahn-Teller distortions [42] in the pairing of second layer Si atoms. DBs which are raised through the pairing of second layer Si atoms exhibit more sp$^3$ qualities creating a more electron-rich state, while DBs which are lowered through the pairing of second layer Si atoms exhibit more sp$^2$ qualities and create a more electron-deficient state [42,69,71]. This resulting lattice shift is predicted to cause an almost ionic charge redistribution throughout the wire via an up-down(sp$^3$-sp$^2$) pairing of DBs within the wire. More recent theoretical studies suggest that spin ordering through electron-electron interactions is the more stable ground state configuration when compared to a charge reordering from electron-phonon interactions [26,27,29,66,72]. Wires are predicted to align in an antiferromagnetic fashion at finite length, with the non-magnetic (ionic) ordering becoming more stable at longer wire lengths due the additional stability of longer ionic configurations [27,29].

Using KPFM, it is revealed that with a net neutral charge, DB wires do not exhibit any charge reordering. Only through the addition of a negative charge (or positive charge as shown in Figure S8) does the DB wire reorder into the theoretically predicted ionic distribution, likely facilitated by an additional polaronic effect [69,73–75], as seen in Figure 3.  Figure 3a shows the Δf(V) spectroscopy taken over the second DB in a five DB wire (indicated by the black arrow in Figure 3 (b)-(f)) revealing four distinct charge distributions, with a Δf(V) line scan map shown in (b). Between -1.0 V and -0.5 V (purple), the DB wire exists in a net neutral charge state with each DB in the wire imaged in neutral charge state, as shown in Figure 3(b)-(c) (insights into the predicted spin ordering require a spin sensitive probe and are currently outside our abilities).  Between -0.5 V and 0.25 V (orange), the DB wire enters a net 1e$^-$ charge state, resulting in an ionic charge redistribution and corresponding lattice distortion (-+-+-, ududu) as shown in Figure 3(b),(d).

As the bias is increased between 0.25 V and 0.45 V (blue), the 5 DB wire is shown to periodically enter an alternative net 1e$^-$ (0-+-0, dudud) charge distribution shown in the top half of Figure 3(e) and in the line scan map between 0.05 V and 0.45 V in Figure 3(b). Such a charge configuration is predicted to have a higher overall energy [27] than the (-+-+-, ududu) configuration and is therefore less stable when imaged. Although the outer atoms in this charge configuration are expected to be in a more electron-depleted sp$^2$ configuration, the rigidity of the end H-Si atoms, which cannot similarly re-hybridize in combination with the crystal doping level, prevent the formation of positive DBs at the ends of the wire [27,29]. Increasing the bias between 0.45 V and 0.8 V (green), the 5 DB wire enters a higher order charge state (multiple e$^-$). Looking at the Δf map of Figure 3(f), the DB wire appears to be in a (-0-0-) charge state, however, Figure 3(b) shows the DB wire appears to be in a (-000-) charge configuration. Since Figure 3(b) was imaged with a 50 pm greater tip-sample separation, the central DB in Figure 3(f) is thought to charge due to an increased attractive interaction between the tip and surface DB (Figure S9 highlights the height dependence of charging within 5 DB line scan maps).

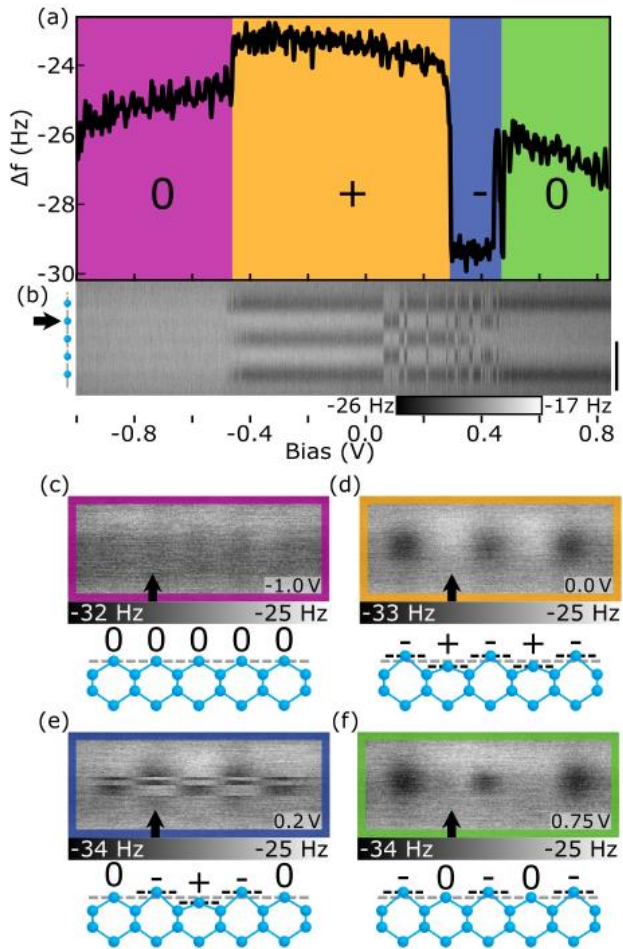

Figure 3. **Charge Distributions of the 5 DB Wire.** (a) Δf(V) spectroscopy over the second Si DB (marked with a black arrow in (b)-(f)) of the 5 DB wire showing four distinct charge regions (Δf(V) spectroscopy of all 5 DBs within the wire is shown in Figure S10). (b) Stacked Δf(V) line scans over the DB wire. 50 line scans at bias increments of 0.02 V from -1.0 to 0.82 V are shown. Scale bar is 1 nm as indicated on the right of (b). (c)-(f) Constant height Δf image of the neutral, lower energy net 1e$^-$, higher energy net 1e$^-$, and higher net negative charge (2e$^-$ or 3e$^-$) configuration, respectively. Each ball and stick model shows the expected trends in lattice

distortion as calculated in Ref [27] and [71]. Bias values for (b)-(e) are -0.7 V, -0.2 V, 0.2 V, and 0.8 V respectively. Height setpoint of (b) is -200 pm and (a),(c)-(f) is -250 pm, as referenced to a common setpoint (see methods). All images are 1.0 x 2.5 nm$^2$.

Extending these measurements to DB wires of varying length, the same charge reordering is seen for DB wires with a net 1e$^-$ charge. Odd DB wires, shown in Figure 4a, are observed in a lower (udu) and higher (dud) energy configuration associated with ionic pairing within the structure. Each of the line profiles in Figure 4 is an averaging of multiple line scans in either configuration taken from full line scan maps in Figure S11 where the local minima in Δf correspond to a negatively charged (u) Si DB and the local maxima in Δf correspond to a positively charged (d) Si DB. Like the 5 DB wire, the 3 and 7 DB wires show a lower energy (-+-, udu) and (-+-+-, udududu) charge reordering, respectively. In the higher energy configuration of the 3 and 7 DB wires, each host Si atom switches its orientation resulting in (0-0, dud) and (0-+-+-0,dududud) configuration respectively. As predicted in Ref. [27], the energy difference between the (udu) and (dud) states of the 3 DB wire is larger than the energy difference between these states in the 5 and 7 DB cases. As a result, the higher energy (0-0, dud) configuration of the 3 DB wire is less stable and was never observed for a complete line scan. The line scan presented in Figure 4(a) only showed a momentary excitation as the tip scans over the centre DB. As a result, the line profile appears much noisier (due to reduced averaging) with each DB appearing negative. The blue line profile was drawn to show the expected configuration if the structure was stable enough to be completely imaged by the tip. The transition of the middle DB in the 3 DB wire from the (+) to (-) charge state is also seen in the Δf(V) spectroscopies shown in Figure S10.

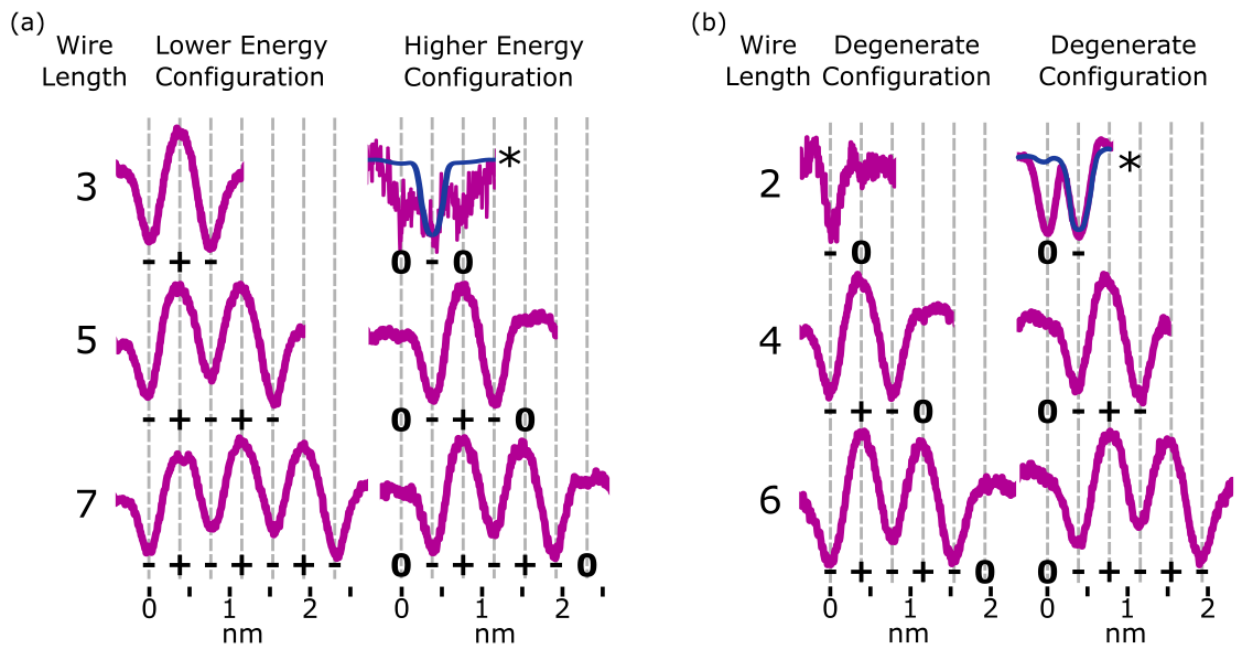

Figure 4. **DB Wire Charge Distributions**. (a),(b) Line profiles of odd and even length DB wires from 2 to 7 DBs in lower energy and higher energy charge configurations (displayed configurations are degenerate for even length DB wires). Each profile shown is an average of 20-50 individual line scans from Figure S11, except the higher energy configuration (0-0) of the 3 DB wire and the degenerate (-0) configuration of the 2 DB wire; these were only visible for 1 and 6 line scans respectively. The blue lines show the expected configuration for each as described in the text.

Unlike the odd length DB wires, the even length DB wires demonstrate mirrored degenerate configurations when in the net 1e$^-$ charge state. Figure 4b shows line profiles of both degenerate distributions for DB wires of length 2, 4, and 6. Much like the odd wire case, DBs which are raised through second layer pairing (sp$^3$) are imaged as negative, while DBs which are lowered through second layer pairing (sp$^2$) are imaged as positive if they sit within the wire and neutral if they sit at the ends of the wire. The ionic distributions in the 2, 4, and 6 DB wires are therefore observed in the (-0/0-, -+-0/0-+-, and -+-+-0/0-+-+-) charge distributions, respectively, as shown in Figure 4b. In the 2DB wire case, the predicted barrier between degenerate configurations is much lower than that of the longer DB wires [42,76], so it toggles more easily between each configuration. As a result, the 2 DB wire configuration often falsely appeared in a (--) configuration, as the negative DB easily "followed" the tip during scanning. The left configuration (-0) appeared for only a few line scans resulting in greater noise seen in the line profile, with the right configuration (0-) never appearing in this dataset, possibly due to the unseen presence of a local electrostatic perturbation from subsurface charges making the negative charge favour a side [17,21]. The blue curve has been drawn to highlight the expected configuration. Full KPFM spectroscopies and Δf(V) line scan maps are shown in Figure S10 and S11 respectively.

As the DB wire length is increased, the bias at which each wire transitions from the net neutral to net negative (1e$^-$) charge state is found to occur at more negative biases, as predicted in Ref [27,29]. Figure 5 highlights this reduction in charge transition voltage for DB wires up to 15 DBs long (extracted from Δf line scan maps in Figure S12). In addition to this associated decrease in charge transition bias with increased length, a slight increase in the charge transition bias is observed between odd length and 1 DB longer even length wires (3 to 4, 5 to 6, etc.). Since these odd and even wires possess the same number of positive and negative DBs, we speculate that this shift in charge transition bias is due to an increase in energy from the additional lattice strain of the end DB without the energy lowering Coulombic contribution from ionic pairing [27,29]. Increasing the wire length further from an even to odd length wire (4 to 5, etc.) creates an additional ionic (+-,du) pair which lowers the voltage at which the charge transition occurs. The opposite trend between odd and even length DB wires was predicted in Ref. [27,29], which we attribute to their models assuming an ionic configuration in the net neutral charge state. At a DB wire length above 9 DBs, this charge transition voltage is found to reach a threshold value of roughly -0.55 V, suggesting the added Coulombic contribution across the wire is screened at these longer lengths. The inset of Figure 5 shows a 12 DB wire in which a domain fracturing behaviour is seen through the decoupling of the two ends which are now able to switch between charge configurations independently. A full discussion of the longer lines is provided with Figures S9 and S10.

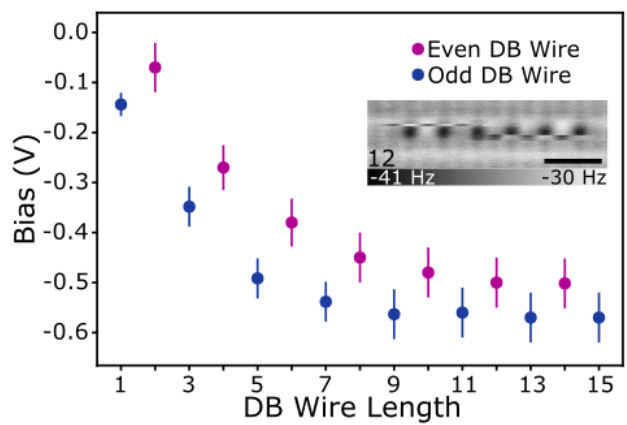

Figure 5. **Charging Voltage of DB wires.** Bias at which the DB wire transitions from a net neutral to net negative charge state (1e-). Error bars account for variations in charging energy between DBs in the Δf(V) spectroscopies (Figure S10) and line scan images (Figure S11 and Figure S13) taken at varying tip heights. The inset shows a 12 DB wire imaged at 0 V and -300 pm to show an example of domain ordering for longer length wires. A full set is provided in Figure S12.

By using a Si terminated tip sensitive to the three different charge states of surface Si DBs, we were able to show the preferred configurations of DB structures at varying probe biases. We were able to confirm the charge redistribution and corresponding buckling of a net neutral bare dimer, as well as identify the pi and pi* states through Δf(V) spectroscopy. Examining DB wires, we report that the theoretically predicted charge redistributions and lattice distortions are only favoured when the entire DB structure is negatively charged. The net neutral charge state of the DB wire reveals no such charge redistribution, allowing for a possible spin ordering to be present. Further measurements using spin sensitive microscopy is needed to confirm this. DB wires in the net negative (1e-) charge state were observed to hold two different configurations, corresponding to a lower (udu) and higher (dud) energy configuration in odd length DB wires and a degenerate (ud/du) configuration in even length DB wires, with the charging voltage dependent on the wire length. Higher order charge configurations were also revealed where additional charging breaks the ionic pairing. These higher order charge distributions likely contribute to the observed character in STM images. These results have allowed for the clear interpretation of charge distributions within DB structures giving insights into their expected viability as charge carriers for atomic scale electronics.

We would like to thank Thomas Dienel, Roshan Achal, Samuel Ng, and Konrad Walus for the valuable discussions regarding these results. We would also like to thank Mark Salomons and Martin Cloutier for their technical expertise. We thank NSERC, Compute Canada, and AITF for support.

JC collected all experimental data except Figure S7 which was collected by TH. The manuscript was written by JC with input from all authors. TH, MR, JW, and EL further supported the development of this manuscript with additional experimental data not shown. JP and RAW supervised the project.

The supporting information features Figures S1-S13 as mentioned in the text, further discussions of the bare dimer, charge distributions within DB wires up to 15 DBs in length, and the experimental methods.

# Supporting Information: Ionic Charge Distributions in Silicon Atomic Wires

Jeremiah Croshaw[1,2], Taleana Huff[3], Mohamad Rashidi[1], John Wood[1], Erika Lloyd[1], Jason Pitters[3], Robert Wolkow[1,2,3]

[1]Department of Physics, University of Alberta, Edmonton, Alberta, T6G 2J1, Canada

[2]Quantum Silicon Inc., Edmonton, Alberta, T6G 2M9, Canada

[3]Nanotechnology Research Centre, National Research Council Canada, Edmonton, Alberta, T6G 2M9, Canada

**Figures S1 to S13 as mentioned in the text**

**Supporting Discussions for the bare dimer, and charge distributions among DB wires.**

**Experimental Methods**

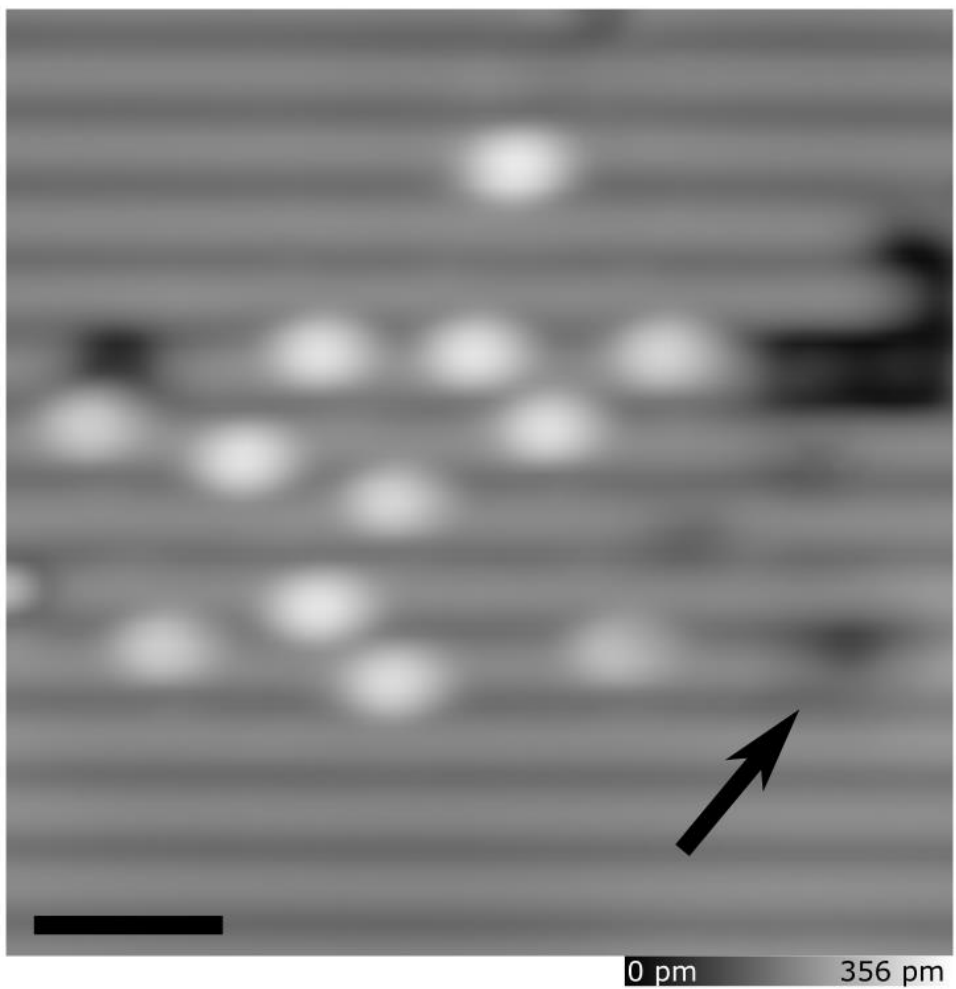

Figure S1. **Cluster of Surface DBs.** Filled states image of the area surrounding the DB from Figure 1 (marked by black arrow). Scale bar is 2 nm.

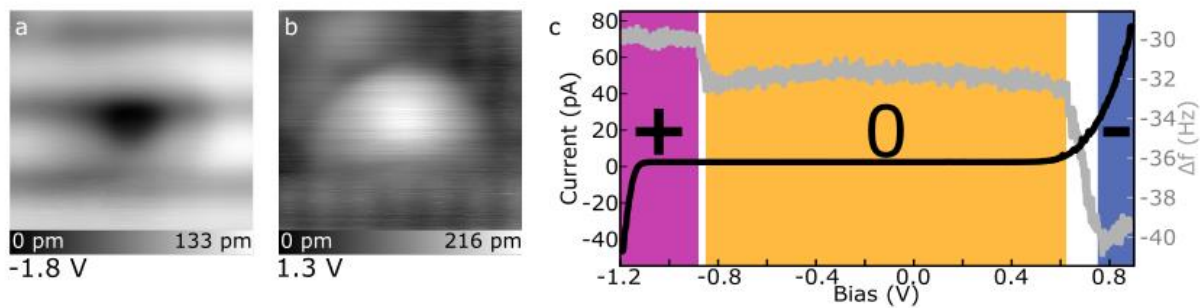

Figure S2. **STM measurements of the DB.** (a) The filled states image of the DB. (b) The empty states image of the DB. Sample biases are indicated in the lower right of each image. Each image is 2.6 x 2.6 nm$^2$ (c) I(V) spectroscopy of the DB compared to the Δf(V) spectroscopy from the main text. The empty and filled states images in (a) and (b) better match the characteristics of a DB on a p-type sample suggesting the nearby electrostatic perturbations have significantly changed modified the emptying and filling rates of the

DB [49,50] rendering it positive at -1.8 V and neutral at 1.3 V as opposed to the expected neutral (-1.8 V) and negative (1.3 V) commonly found on n-type samples [47].

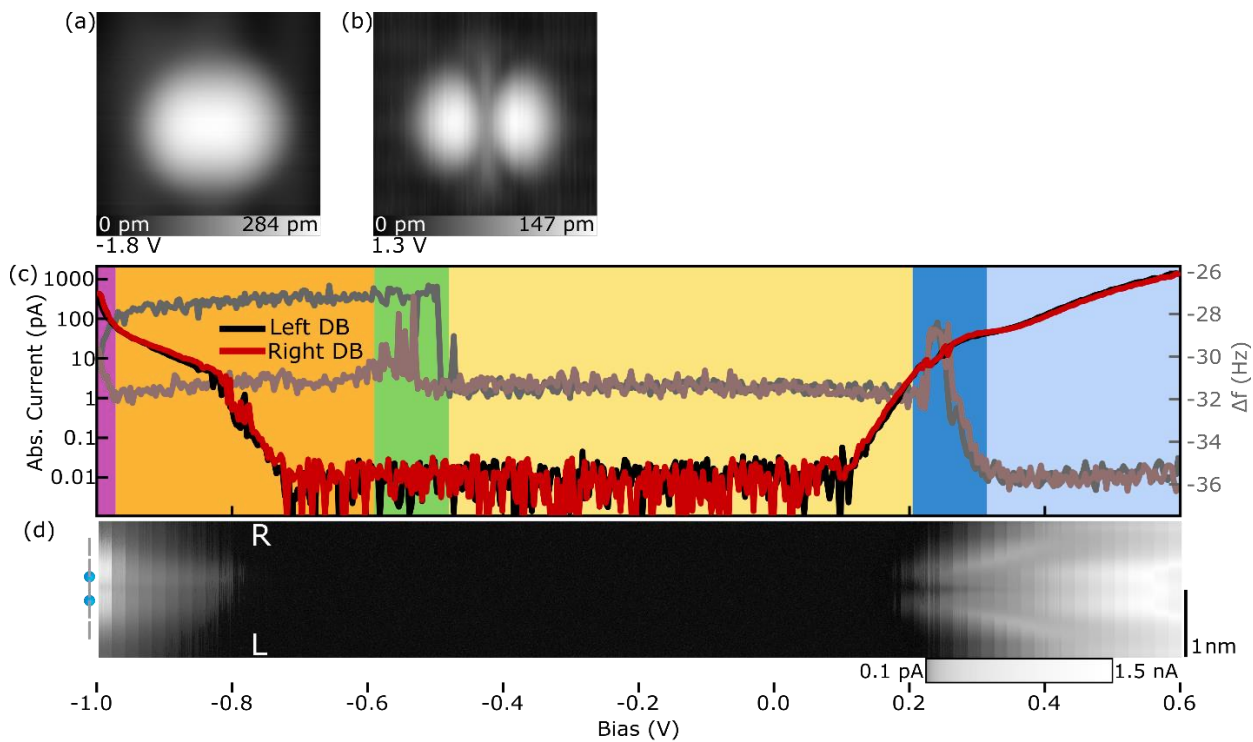

Figure S3. **Bare Dimer Height and Δf maps.** (a) filled (-1.8 V) and (b) empty (1.3 V) states STM image of the bare dimer structure.(c) I(V) spectroscopy of the bare dimer compared to the Δf(V) spectroscopy from the main text. (f) I(V) line scan maps simultaneously measured with the Δf(V) line scan maps in Figure 2f. The position of the bare dimer in the line scans is indicated on the left of (d) with the grey line added to orient the structure. 50 line scans at bias increments of 0.02 V are shown. The current scale in (c) and (d) is plotted logarithmically to highlight low current features at the onset of the conduction and valence band. Relative heights for each are -300 pm. The scale bar for the line scans is 1 nm as indicated on the right (d).

**The Bare Dimer**

The bare dimer offers a unique structural formation compared to other DB structures of the H-Si(100) surface due to the presence of a dimer bond. As outlined in the main text, this unique structural property results in six discernable regions within the probed bias range. By looking at the Δf(V) spectroscopies over both DBs in both forward and backward scan directions, and at various heights as shown in Figure S4, further insights into the structural and charge states of the bare dimer can be gained. The left panel shows the Δf(V) spectroscopies over the left DB with the forward direction indicating a bias sweep from -1.0 V to 0.6 V (0.6 V to -1.0 V for the backward direction) while the right panel shows spectroscopies over the right DB. Each spectrum is taken between -200 pm and -350 pm with 50 pm increments. The spectra with more negative Δf signals correspond to a reduced tip-sample separation due to an increased coulombic interaction between the tip and sample. The spectroscopy features corresponding to each of the six bias windows are visible for all heights with a slight variation in their onset due to the variation in tip induced band bending with changing heights. One of the key differences seen between the left and right DBs occur in the orange, green, and yellow regions. In the yellow region, there is a hysteresis seen between the forward and backward scans of the left DB that is not seen in the right DB. As the tip probes the left DB within the yellow region in the forward direction, the DB transitions from positive to negative on the time order of a few seconds. Scanning in the backward direction, the left DB stays in the negative charge state until the bias reaches values within the green and orange region. The length at which the left DB stays in the positive state before transitioning to the negative state in the yellow region decreases with reduced tip height suggesting it is influenced by a stronger interaction with the tip. With the exception of a short lived transition to the positive charge state at -250 pm, the right DB does not show this hysteresis between the forward and backward directions, remaining instead in the negative charge state.

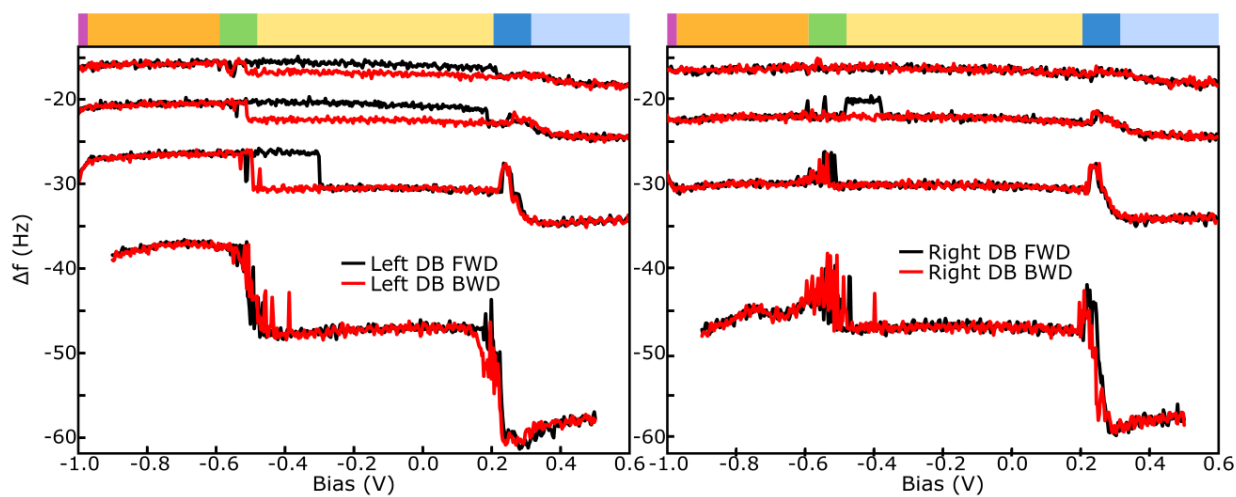

Figure S4 **The Bare Dimer Δf(V) Spectroscopies.** Δf(V) spectroscopy over the left and right DB as indicated. The black curves show the forward scan direction (-1.0 V to 0.6 V) while the red curves show the backward direction (0.6 V to -1.0 V). Each spectroscopy over the left and right DB are taken at different heights corresponding to a tip offset of -200 pm (less negative Δf), -250 pm, -300 pm, and -350 pm (more negative Δf). The bias regions mentioned in the text are coloured above each panel.

To better understand the influence of the tip within this bias region, it is useful to consider the electron energy levels of the left and right DB as a function of the nuclear coordinates of the host Si atom relative to the surface [77–79]. Figure S5 (a) shows the double potential well of the two DBs in a bare dimer. In this instance, the left DB (drawn in red) is shown in the positive charge state with the more $sp^2$-like character of the host Si atom indicated by the leftward shift along the horizontal axis. The right DB (black) is shown in the negative charge state as indicated by the presence of two electrons (black arrows) with the more $sp^3$-like character represented by the rightward shift in the nuclear coordinate. The horizontal dashed lines within the potential indicate possible electron levels available through an electron-phonon coupling [78,79]. The observed charge state of each DB is facilitated by the emptying and filling rates between each DB and the conduction band, the DBs and the tip, and between the DBs of the bare dimer as indicated with the grey arrows. The coloured regions are the same as the main text and are used to reference the position of the tip Fermi level.

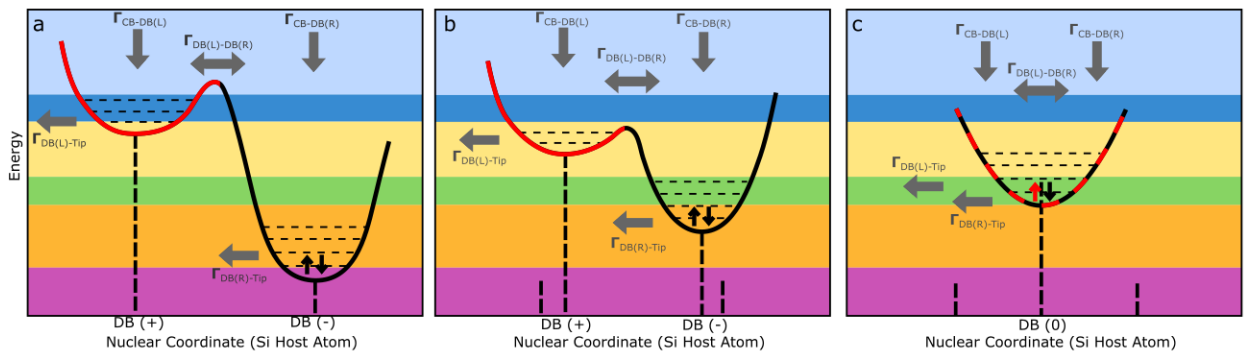

Figure S5 **The Bare Dimer Charge States with Atom position.** (a) Qualitative double potential well system of the buckled bare dimer showing the left DB (red) in the positive charge state and the right DB (black) in the negative charge state. The difference in charge state is accompanied by different nuclear coordinates of the host silicon atom indicated along the horizontal axis. The coloured regions match those presented in the spectroscopies of the main text and SI figures which are used to reference the position of the tip Fermi level. The grey arrows indicate potential current pathways between the conduction band and the DBs, the DBs and the tip, and between DBs. The horizontal lines within the potential well indicate possible electron energy levels made available through electron-phonon coupling. (b) The same potential well in (a) except an attractive interaction between the tip and left DB has shifted the nuclear coordinates of the host Si atom corresponding to a shift in energy level for each DB. The original lattice position is shown by the two shorter vertical dashed lines. (c) The double potential well system of the symmetric bare dimer where both DBs now sit at the same relative Nuclear Coordinate resulting in the energetic alignment of the two energy levels.

To appreciate the dynamics within the yellow region, consider a tip positioned above the left (positive) DB. The attractive interaction from the tip raises [18,55,59] the host Si atom from the surface changing the nuclear coordinate in both the left and right DB as indicated in Figure S5 (b). With a reduced separation in energy between the filled, negative DB and the empty, positive DB, it becomes possible for an electron to thermally excite to the left DB leaving the bare dimer in an unbuckled, symmetric configuration shown in Figure S5 (c). Following this excitation, the bare dimer then buckles again to the ionic (-+) configuration. Due to the interaction of the tip, the left DB now preferentially buckles to host the negative charge. If the tip is positioned over the negative DB in the bare dimer (right DB or the left DB in the backward direction), the charge transition from a negative to positive DB does not occur. The additional interaction from the tip attracts the DB, keeping it in a preferred $sp^3$ configuration, preventing an electron from transferring to the positive charge state which would unbuckle the dimer. As mentioned in the main text, the preferred configuration of the buckled bare dimer is likely dictated by some unseen electrostatic perturbation in the subsurface region. It is important to emphasize that during this transition, the Fermi level of the tip never drops below the energy level of the negative DB and so all the electron dynamics within the bare dimer are solely due to the movement of charge within the bare dimer. Additionally, if the tip position is allowed to laterally move from the negative DB to the positive DB within the bare dimer, (as in the Δf line scan maps), it allows for the bare dimer to buckle between the two degenerate configurations. The likelihood of a transition is dictated by tip height with larger tip sample separations resulting in fewer transitions and smaller tip sample separations resulting in more transitions as shown in Figure S6.

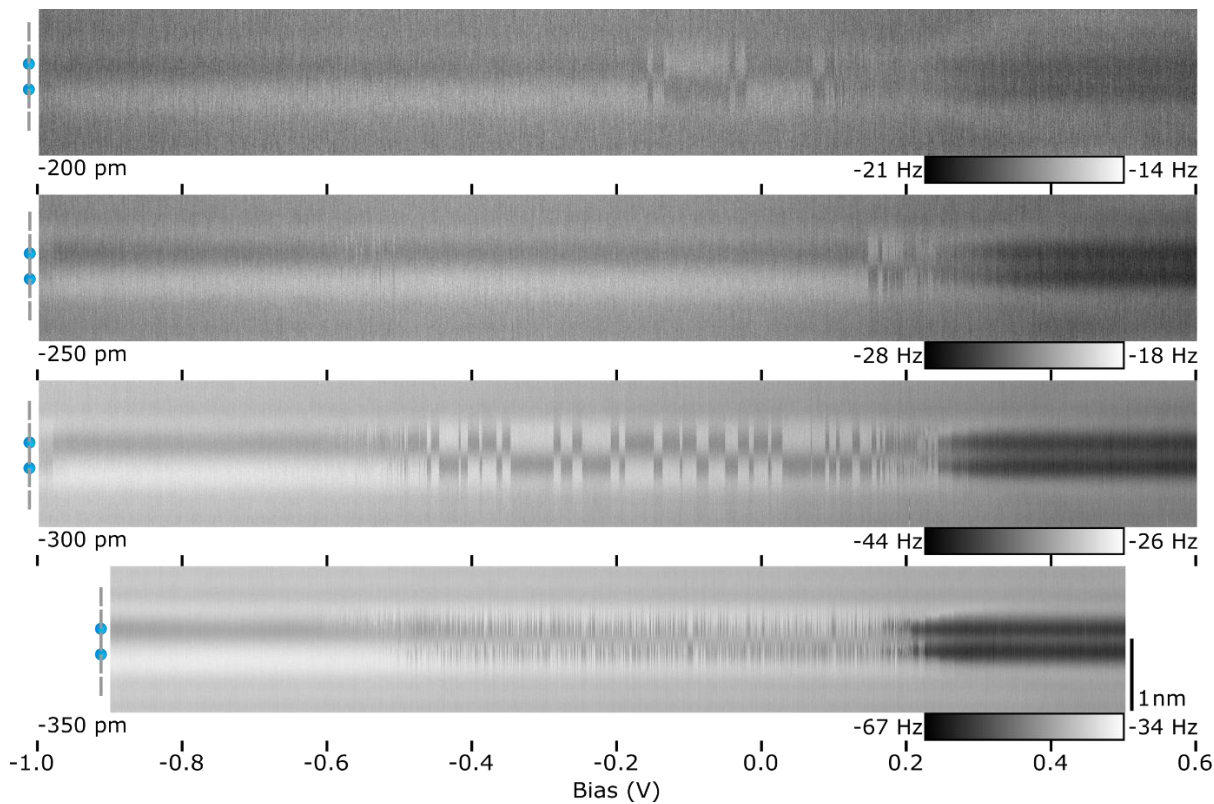

Figure S6 **The Bare Dimer Δf(V) line scan maps at various tip heights.** Imaging parameters are the same as Figure 2. The relative tip height is indicated in the lower left corner of each line scan map.

Looking at the green region of Figure S4, the rapid switching in Δf can be seen in both the left and right DB in the forward and backward directions suggesting that the dynamics seen are not dependent on tip position over the bare dimer. We attribute this rapid switching between states to the resonance of the unbuckled pi state of the bare dimer with the Fermi level of the tip. The dynamics are then dictated by the competing emptying rates from the DB to the tip ($\Gamma_{DB(R/L)-Tip}$) and the filling rates from the conduction band to the DB ($\Gamma_{CB-DB(R/L)}$) as well as the charge transfer between the DBs ($\Gamma_{DB(R/L)-DB(L\backslash R)}$), and corresponding buckling and unbuckling associated with charge localization within the bare dimer. It is currently unknown what mechanism drives the bare dimer into the symmetric configuration since such signals are seen over both the positive and negative DB. Correlating the reduction in frequency of such switching signals to increased tip sample separations, it is possible the bare dimer dynamics could be facilitated by the oscillation of the AFM tip, however, further studies are needed to support this claim.

Lastly, the orange region shows the absence of switching between charge states with the left DB imaging as positive and the right DB imaging as negative in both the forward and backward direction. Looking at Figure S5 (a), it can be seen that the Fermi level of the tip is now between the proposed electron energy levels of the symmetric bare dimer and the electron levels of the fully buckled negative DB.  Much like the yellow region, interactions with the tip would change the electron energy level of the negative DB due to tip induced variations to the nuclear coordinate. The difference is that within the orange region, the electrons within the negative DB would be emptied to the tip before the bare dimer reaches its symmetric configuration.  If the subsequent filling rate from the conduction band is quicker than any lattice distortion and charge redistribution within the bare dimer, it follows that no switching of the bare dimer configuration would occur.  This claim is supported by the observation of current at roughly -0.75 V in Figure S3 (c) with no modification to the charge state in the Δf signal seen in the background of Figure S3 (c). Investigating the dynamics of such emptying and filling rates are outside the abilities of conventional STM and nc-AFM and warrant further investigation using time-resolved techniques.

Insights into the purple, dark blue, and light blue region can also be gained from looking at Figure S5.  In the purple region, the Fermi level of the tip is now below the energy level of the fully buckled negative DB allowing for a constant emptying of this state.  Although the constant height image in the bias range shows the bare dimer in a symmetric configuration, the proposed energy alignment of the bare dimer with the Fermi level of the tip suggests that it sits in a (0+) configuration that may buckle between the (0+)/(+0) configurations at a rate faster than the AFM sampling rate giving the symmetric appearance in the Δf images.  The dark blue and light blue regions now bring the energy level of the positive DB in alignment with the Fermi level of the tip indicating that this electron level may now be filled by the tip. The unique features seen in the Δf(V) spectroscopies and line scans maps corresponding to the sharp increase in Δf and sudden decrease following the transition from the dark blue to the light blue region suggest a combination of lattice and charge dynamics outside the sampling rates of our STM and AFM and warrant subsequent investigation. As mentioned in the main text, the more negative shift in Δf likely corresponds to a net charging of the DB structure, but charge dynamics make it difficult to assess if the DB structure is rapidly switching between a net 1e$^-$ (0-)(-0) charge configuration or sits in a 2e$^-$ (--) configuration.

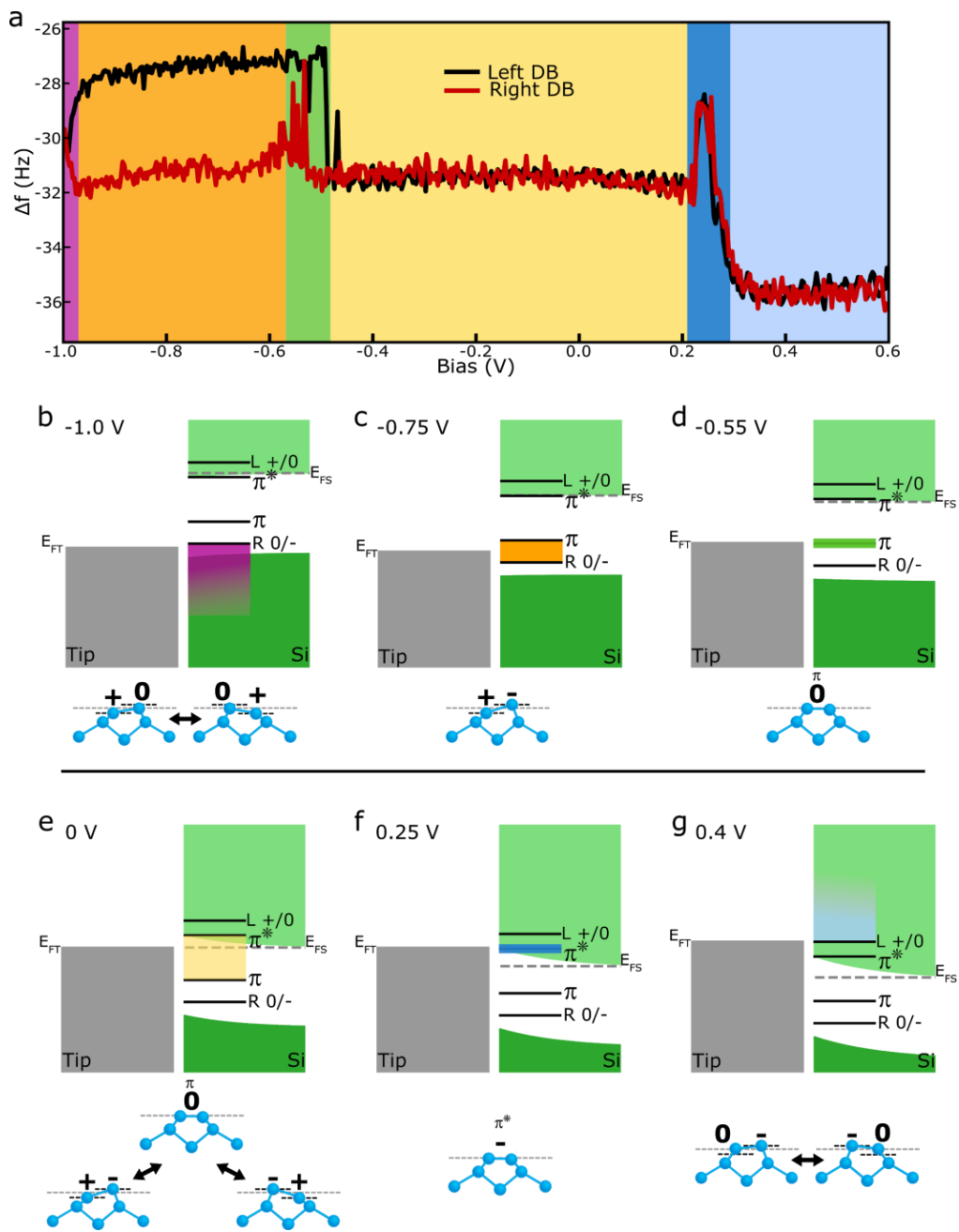

Figure S7. **Band Diagrams for the Associated Δf(V) Spectrospies of a Bare Dimer**. (a) The Δf(V) spectroscopy of a bare dimer as shown in Figure 2 of the main text. (b)-(g) The associated predicted band diagrams of the bare dimer system at bias values indicated in the top left of each panel. The bulk Si band diagrams were calculated using semi-tip [80]. The ball and stick images show the expected lattice and charge distribution as discussed with Figure S5. It is important to note that the pi and pi* states drawn in the band diagrams are only available when the bare dimer is in a symmetric configuration and the L +/0 and R 0/- charge transition levels only exist when the bare dimer in the buckled configuration.

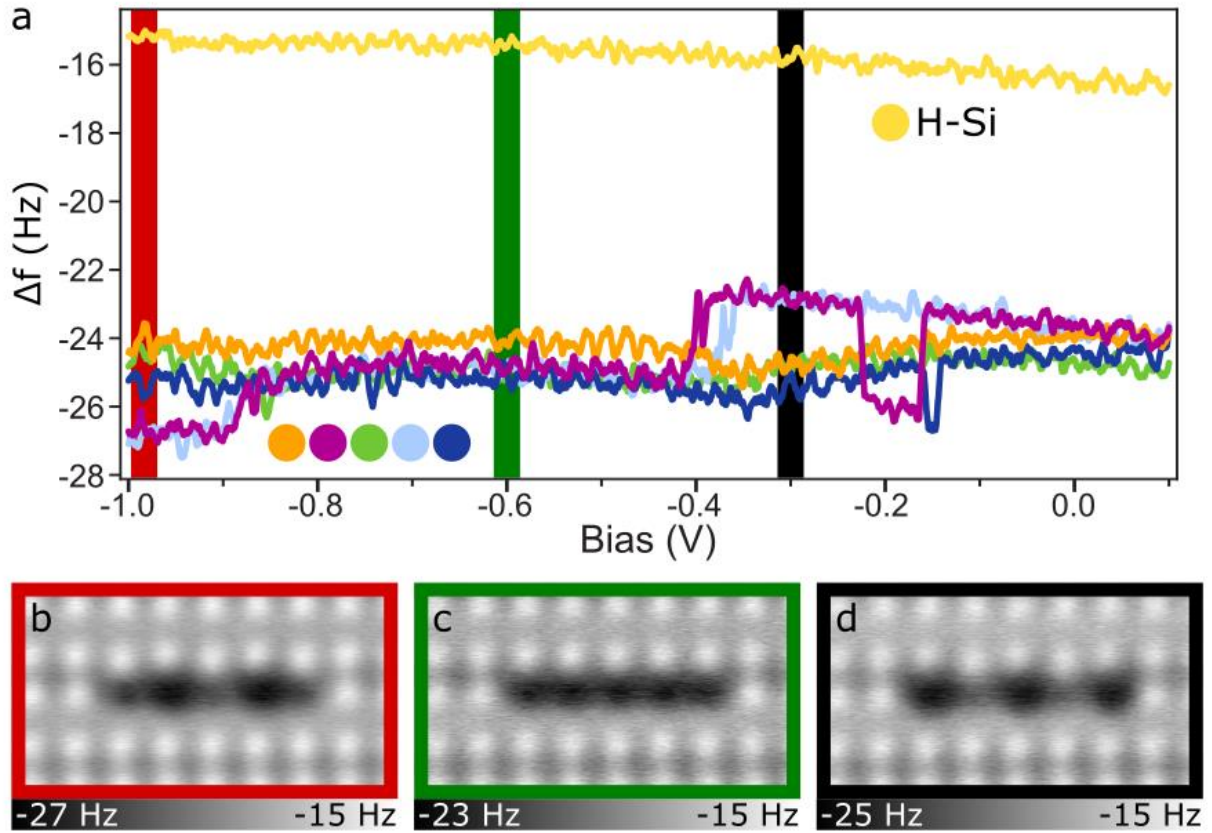

Figure S8: **Net Positive Configuration of a 5 DB wire as probed with an H terminated tip.**
(a) Δf(V) spectra taken over each DB in the 5 DB wire as indicated in the lower left, as well as over the hydrogen surface in yellow (tip height = −280 pm). A Savitsky-Golay filter of order 9 was applied to allow easier differentiation of the curves. (b-d) Constant-height Δf AFM images of the wire with bias values as indicated in (a). Each image is 1.9 x 3.4 $nm^2$ with a relative tip height of -310 pm. The presence of a net positive lattice distribution in DB wires was not as easily observed for results presented in the main text due to an increase in current from the valence band at the onset of the net positive charge transition associated with a different alignment of the DB energies with the Fermi level of the crystal from varying sample preparations [81]. Charge transitions correlated to the net positive distribution can be seen for the 5 and 7 DB case in Figure S11 but are short lived due to the onset of sustained tunneling current (not shown). Further studies on p-type crystals using a Si terminated AFM tip would be useful to better demonstrate the net positive redistribution of DB wires.

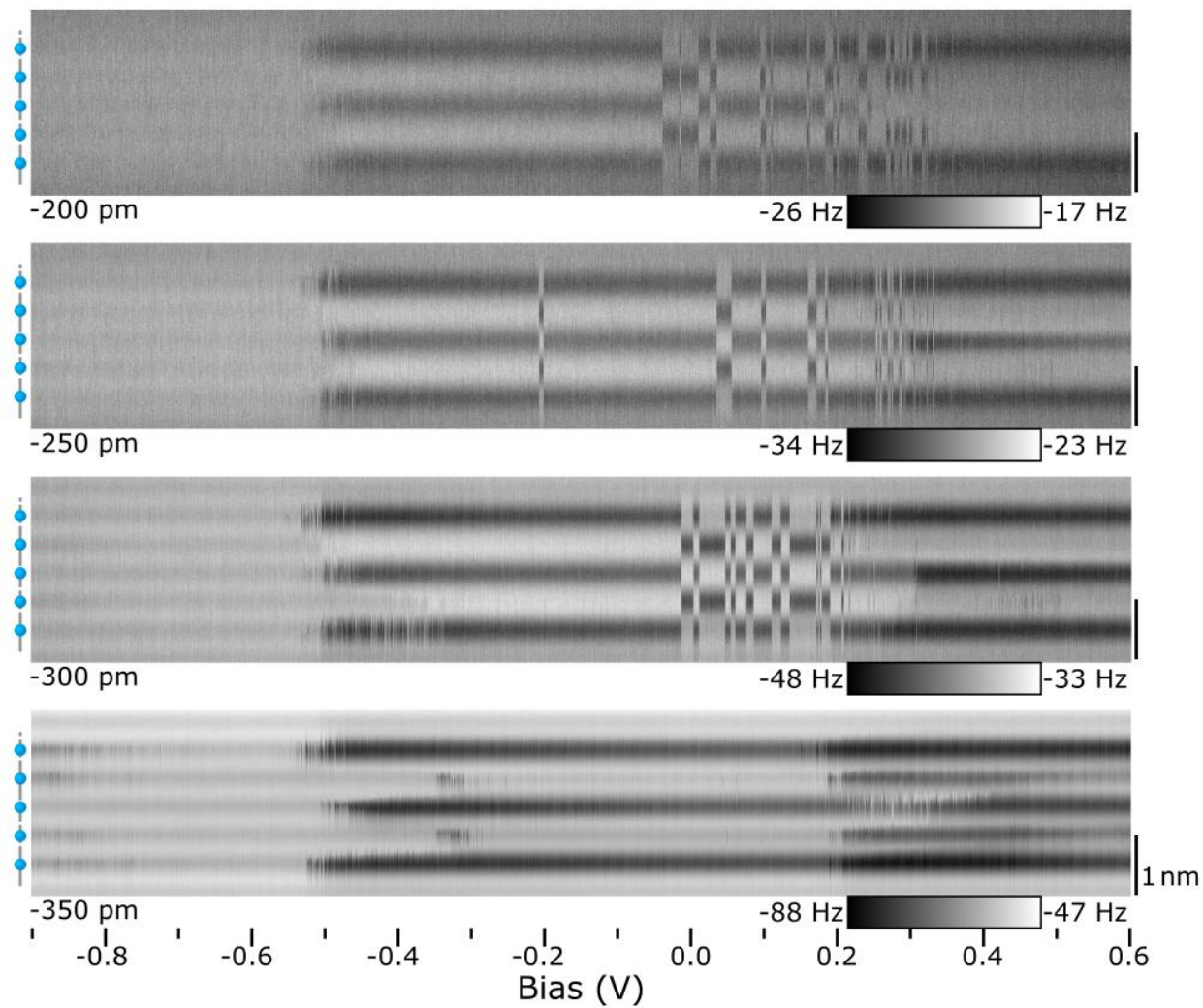

Figure S9. **Charge Distribution with Tip Height.** Line scans of a 5 DB wire. Bias increments are 0.02 V with 50 line scans taken at each bias. Each image is taken at a different relative depth as indicated in the bottom left of each set of scans. The streakiness of the centre DB at -350 pm is dependent on scan direction which in this case is top to bottom.

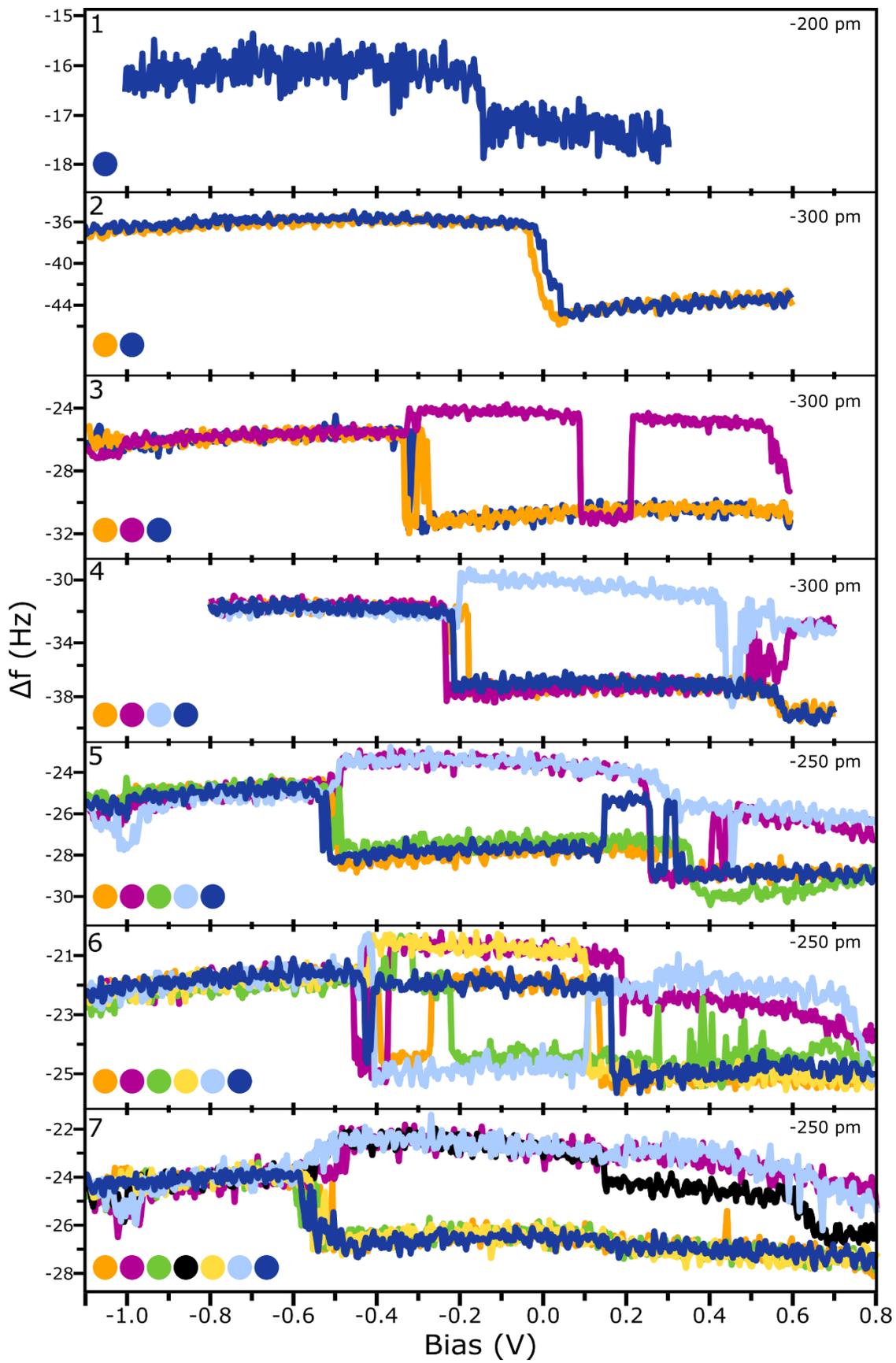

Figure S10. **KPFM Spectroscopy of DB Wires.** Δf(V) spectroscopy over DB wires length 1 to 7 as indicated in the top left of each panel. Spectroscopy colors are shown in the bottom left of each panel. The height of each spectroscopy is indicated in the top right and was chosen to highlight the charge switching of each DB without significant influence from the tip. Note that the centre DB of the 3 DB wire briefly enters a (-) charge state as mentioned in the text. Since neither the left or right DB enter the expected neutral state, the middle DB is thought to only enter this higher energy state due to the sustained interaction with the tip as discussed in the main text.

**Charging Biases and DB wires of Greater Length**

As discussed in the main text, the energy at which each wire transitions from the net neutral to net negative ($1e^-$) charge state depends on the wire length with longer wires ionizing at lower sample biases as predicted in Ref [27,29]. Figure S11 highlights the first ionization bias for DB wires length 2 to 7, along with the second ionization bias where a wire can facilitate an additional negative charge. The 2 DB wire undergoes the first ionization at the largest sample bias around -0.08 V while the 7 DB wire is ionized at -0.5 V. A similar trend can be seen when looking at the secondary ionization energy transitioning to a multiple negative charge state in DB wires as shown in Figure S11. As the sample bias increases and the DB wire is forced to accept an additional charge from the tip, the DB wires redistribute their charge so that the separation between negative charges is maximized. This is most easily observed for the 4 and 5 DB wire case which now contain two negative DBs located at the wire edges with all middle DBs imaged in the neutral charge state. The 6 and 7 DB wires also readjust to hold a negative charge in both wire ends. The 7 DB wire easily accommodates the addition of the negative charge by neutralizing the centre DB leaving two (-+-, dud) clusters at the ends of the wire. The 6 DB wire appears to be in a more frustrated configuration with both outer DBs appearing negative and the two inner DBs showing a switching behaviour similar to the 2 DB case. It is likely that this DB cluster is switching between a degenerate (-+-00-/-00-+-) charge distribution as the tip scans over the wire although confirmation of such a configuration is outside the available sampling rates.

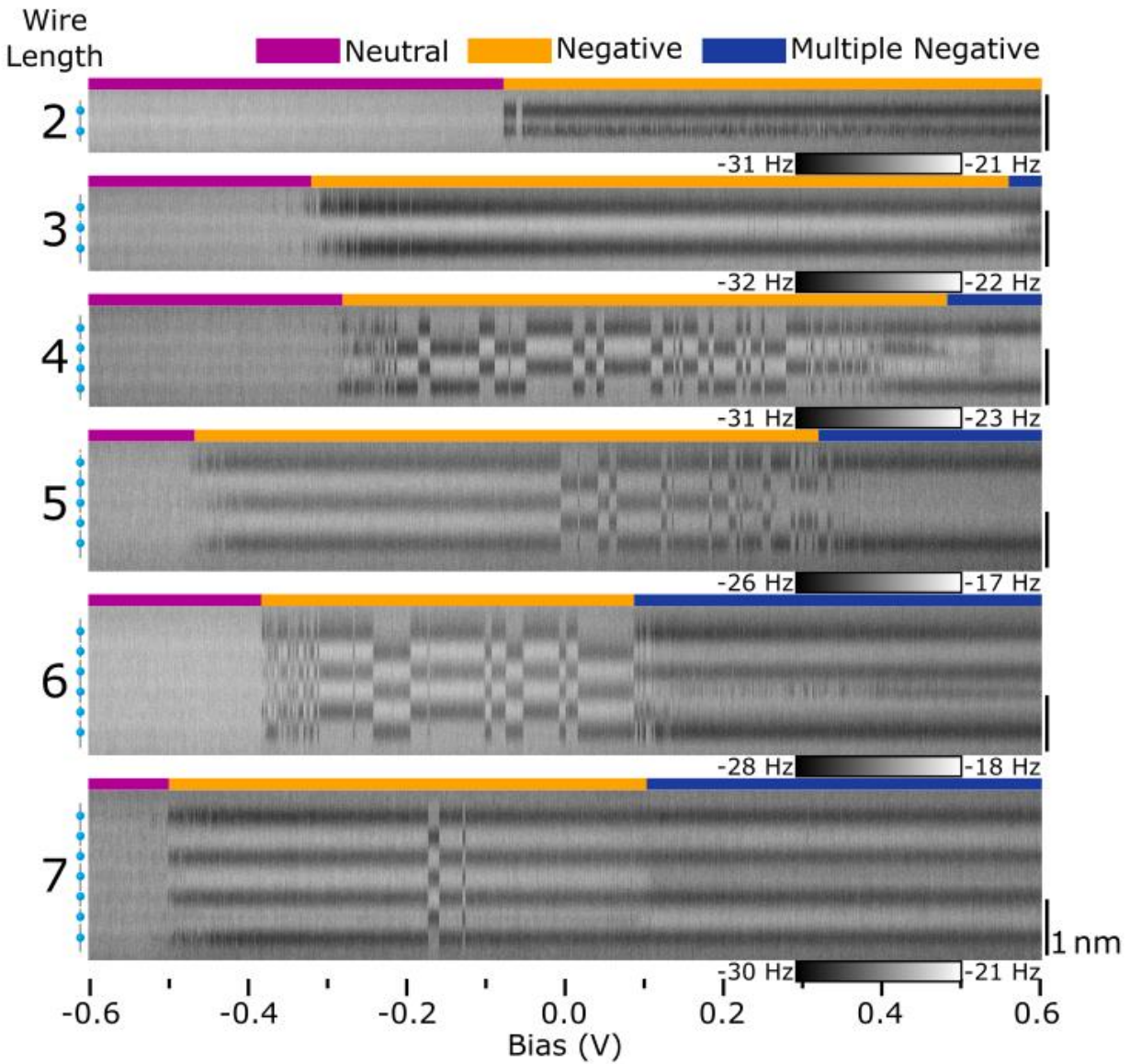

Figure S11. **Charging of DB wires.** Line scans of DB wires length 2 to 7. Bias increments are 0.02 V with 50 line scans taken at each bias. Purple indicates the bias window when the DB wire is neutral, yellow indicates a single negative charge in either the lower or higher order (odd) or degenerate (even) configuration, and blue indicates an observed charge greater than -1. The position of each DB in the DB wires are indicated to the left of each line scan in blue. Line scans taken for DB wires of length 2,4,6 and 7 were taken at -250 pm, 3 was taken at -300 pm, 5 was taken at -200 pm. These heights were selected for each wire to minimize the effects of tip induced switching. The onset of the first ionization event did not vary significantly with decreased tip height as shown in Figure S9.

As the DB length is increased, the DB wires can more easily accommodate a greater number of charges. Since this added charge introduces an additional polaronic effect [69,73–75], the DB wires are able to reorder into a greater number of configurations which are stabilized by this added charge. Figure

S12 shows select Δf maps of DB wires length 8 to 15 at 0 V.  At this bias, the 8 DB wire has transitioned from the (ud) configuration expected for even wires and instead features two (-+-) clusters at either end. The (ud/du) configuration for the 8 DB wire can be seen in the line scans of Figure S13 at bias values lower than the configuration seen in Figure S12 (a), suggesting that the configuration shown in Figure 12 (a) is a higher energy configuration.  The 9 DB wire of Figure S12 (b) is shown to switch between two configurations mid-scan.  The position of the (-+-) cluster on the left side of the 9 DB wire shifts one lattice site with the middle DB appearing negative only when the (-+-) cluster is at the end of the DB wire.  Its appearance is very similar to the 5 DB wire at reduced tip sample separation suggesting this DB may only be negative due to a tip interaction.  The right side of the 9 DB structure does not appear to switch with the left side of the wire which supports the claim that the added polaronic effect from additional charge can stabilize the wires into higher energy configurations, allowing for the wires to decoupling into smaller subunits.  The 10 DB wire again features (-+-) clusters at the edges of the wire with the central two DBs appearing negative. The line scans in Figure S13 again reveal that this is a higher energy configuration compared to the (ud/du) configuration expected with even length DB wires. The shape of the two central DBs appears much more asymmetric compared to all other negative DBs seen suggesting that this is likely a single negative DB which switches between the middle left and middle right DB due to a tip induced charging (like the 2 and 4 DB wire cases), resulting in a net charge of $3e^-$ for the 10 DB wire at 0 V.  The 11 DB wire shows both (-+-) clusters at the ends of the wires with 2 negative DBs separated by a neutral DB giving the structure an apparent charge of $4e^-$.

      The DBs of length 12 to 15 show similar trends as observed in DB wires of previous lengths. The 12 and 15 DB wires both show a switching of the charge configuration within the wire with a decoupling of the left and right sides.  The 13 and 14 DB wires appear to be in a stable configuration with the centre DB imaging very similar to that of the 5 and 9 DB wire suggesting it is negative due to a tip induced charging.  The full line scans for the 8 to 15 DB wires presented in Figure S13 clearly indicate that an increased wire length allows for a greater degree of charge distributions within the wires. The increased length means that charges are more effectively screened along the wires resulting in a decoupling of the wire into smaller subunits.  As the tip bias is increased, the structure of the DB wire can reconfigure due to the stabilizing polaronic effect from charge injection from the tip. Since the DB wires appear to more easily switch between the various configurations, it makes it difficult to assign a single lattice configuration as was done for DB wires of shorter length.

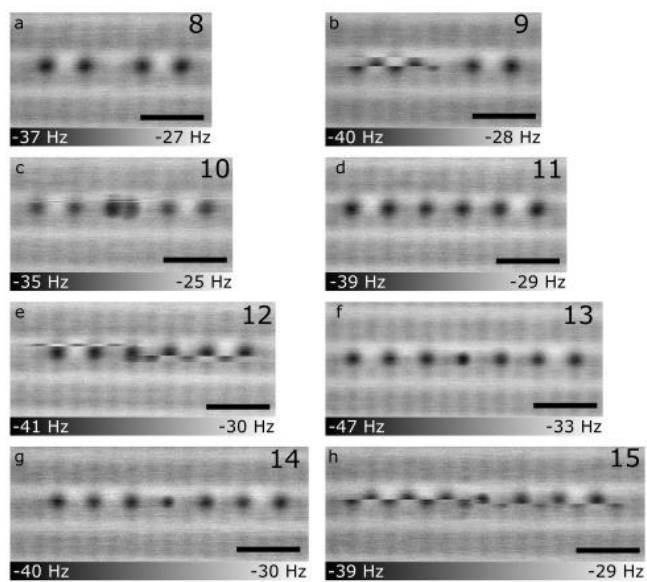

Figure S12. **DB wires of Longer Length**. (a)-(h) Δf(V) maps of DB wires of length 8 to 15 as indicated in the top right of each image. All images are taken at 0 V and -300 pm. Scale bar is 1nm.

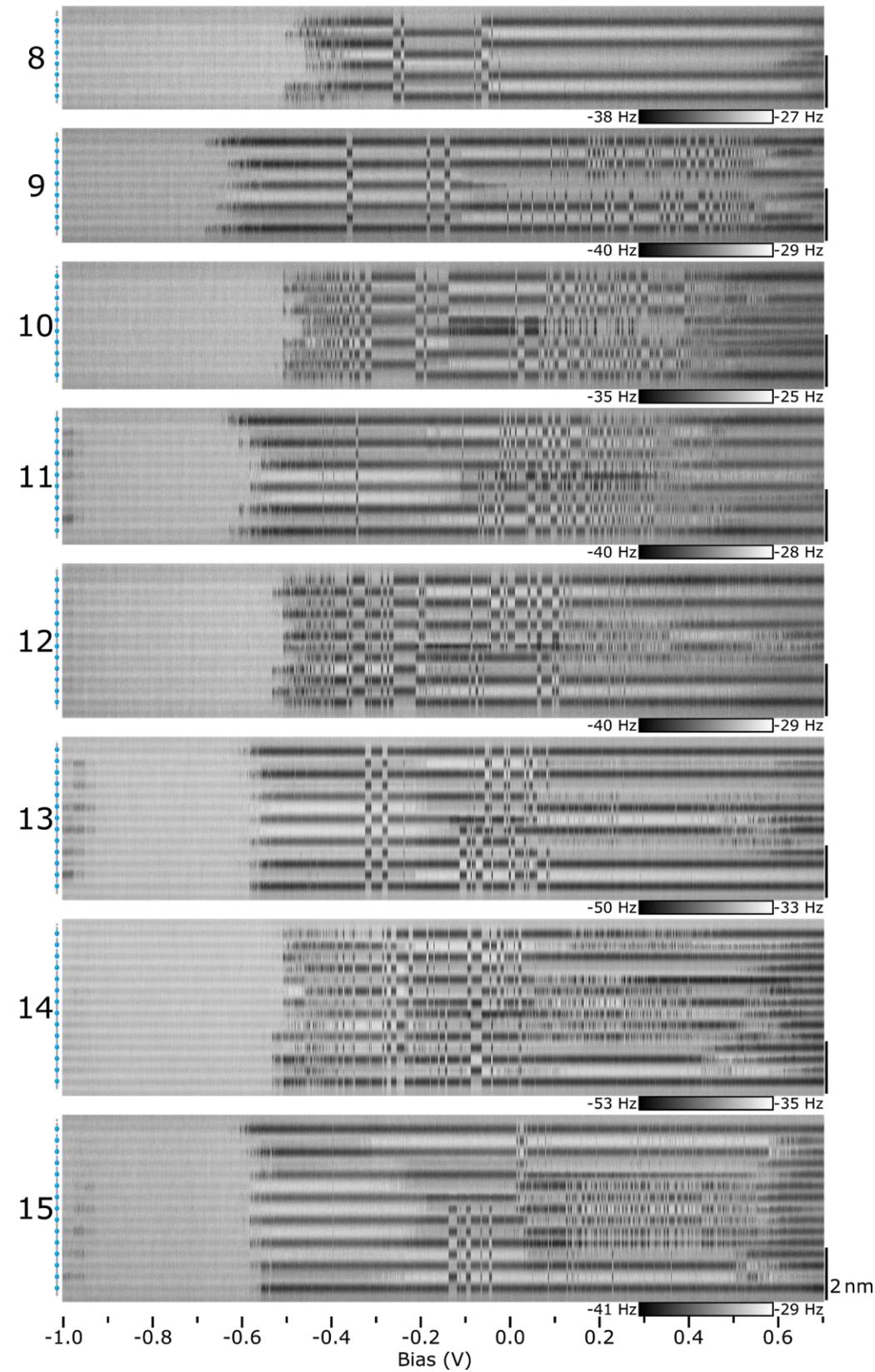

Figure S13. **Charge Distribution Among Longer Wires.** Line scans of DB wires length 8 to 15. Bias increments are 0.02 V with 50 line scans taken at each bias. All line scans were taken at a relative height of -300 pm. The position of DBs within the DB wires is indicated to the left of the line scans in blue.

**Methods**

Experiments were performed using an Omicron qPlus LT AFM [82,83] and an Omicron LT STM system operating at 4.5 K and 3 x$10^{-11}$ Torr. STM tips were fabricated from polycrystalline tungsten wire by electrochemical etching followed by resistive heating and field-evaporation to clean and sharpen the tip using a field ion microscope (FIM) [84]. AFM tips used a third generation Giessibl tuning fork with a focused ion beam (FIB) mounted Tungsten tip ($f_0$ ~ 28 kHz, Q-factor ~ 16k-22k, Amplitude = 50 pm) [85]. The AFM tip was cleaned and sharpened in-situ via controlled tip contacts with hydrogen desorbed patches of Silicon until it returned characteristics corresponding to a Si terminated tip [25,46]. DB structures were created via controlled bias pulses [20,23,86]. The bias ranges for each set of measurements were chosen to capture as great a bias window as possible while trying to prevent any unwanted tip changes due to high tunneling current through the valence and conduction band of the crystal. The AFM measurements taken in this work were taken over several months, so while each tip shows the desired Si tip contrast of the surface, the exact reactivity of the tip varies between some data sets.

Samples were degenerately arsenic-doped (1.5x$10^{19}$ cm$^{-3}$). Samples were degassed at 600 °C overnight followed by multiple annealing flashes at 1250 °C. The samples were subsequently hydrogen terminated by exposing the system to molecular hydrogen (1 x $10^{-6}$ Torr) for two minutes while holding the sample at 330 °C. A tungsten filament held at 1,600 °C was used to crack the hydrogen [48,81].

Images and data were collected using a Nanonis SPM controller, with additional custom-built LabVIEW controllers. The height setpoint references in constant height measurements are relative to a tip-sample separation distance initialized with a tunneling current feedback of 50 pA and sample bias of -1.8 V with the tip positioned over a H-Si atom.